# Age-dependent distribution of officially reported cases of vector-borne infections

*Francisco Antônio Bezerra Coutinho[1,2], Marcos Amaku[1], Esper Georges Kallas[1,3] and Eduardo Massad[3,4,*]*

1. Faculdade de Medicina, Universidade de São Paulo
2. Instituto de Física, Universidade de São Paulo
3. Instituto Butantan
4. Faculdade de Saúde Pública, Universidade de São Paulo
   *corresponding author: edmassad@dim.fm.usp.br

**ABSTRACT**

**OBJECTIVE:** To propose a new approach to analyze the age-distribution of reported cases for vector-transmitted infections.

**METHODS:** Using officially reported number of cases of dengue, Zika, chikungunya, malaria and leishmaniasis for distinct geographical areas, in different periods. Data were treated in special but well-known procedure, transforming the raw data into a density age-dependent distribution and fitting a special continuous function to it.

**RESULTS:** We found that the proportion of age-dependent cases with respect to the total number of cases in a given year (or any transmission season) is probably determined by the ecological interactions between vectors and hosts. The age-distribution of the proportion of cases for the three Aedes-related infections are essentially the same independently of the magnitude of the outbreak and the geographical region considered. On the other hand, for the infections transmitted by other vectors, the age-distributions of the proportion of cases are entirely different.

**CONCLUSIONS:** During specific outbreaks, the ratio between the age distribution of the proportion of officially reported cases and the total number of cases for Aedes transmitted infections such as dengue, chikungunya and zika is **independent** of the size of the outbreak, the size of the studied population, the period when the outbreak occurs; and the geographical region considered. Our results also suggest that the age-distribution of cases is mainly due to the interaction between vectors and their hosts.

Since the interaction between the vectors and male and female hosts may be different, we investigate this and, in fact, the age-distribution is slightly different.

**DESCRIPTORS**. Dengue, Zika, Chikungunya, Malaria, Leishmaniasis, Age-dependent distribution of cases.

**INTRODUCTION**

Vector-borne infections frequently display strong age- and time-dependent patterns in incidence, prevalence, and disease severity[1]. Understanding the age distribution of cases is central to epidemiological inference, estimation of transmission parameters, such as the average age of infection (also referred to in the literature as the average age of prime infection[2]), and more importantly, in the design of targeted control strategies (work in preparation).

In this paper, we consider the following diseases: one transmitted by *Anopheles* mosquitoes (malaria), one transmitted by phlebotomine sandfly (visceral leishmaniasis) and three transmitted by *Aedes* mosquitoes in urban areas (dengue, zika and chikungunya). We focus on the age distribution of infection in the human hosts during a given outbreak. Our main hypothesis is that the age distribution of cases is primarily dependent by the contact patterns between vectors and hosts.

Because *Aedes* mosquitoes transmit dengue, Zika and chikungunya, if the above hypothesis is correct, one would expect that the age distribution of cases for these three infection to be similar. We demonstrate that this pattern is indeed observed empirically (see Figure (10) below). In contrast, infections transmitted by different vectors exhibit markedly different age distributions, which is also supported by empirical observations (see Figure (11) below).

Daily activity biting habits of vectors, occupational behavior of hosts, and housing conditions[3,4], influence human-vector contact. Differences in vector behavior—such as indoor versus outdoor biting—may further shape the age distribution of cases[5]. In this paper, however, we analyze the age distribution of Aedes-transmitted infections in large and medium-size urban centers and for Brazil as a whole. In Brazilian urban centers the transmission is mainly peri-domestic[6,7]. Furthermore, in these settings, the infection

'travels' throughout the city carried by infected hosts[8]. This is clearly illustrated by dengue transmission in São Paulo city[8] (approximately 12 million inhabitants distributed among 1,700 neighborhoods).

For infections transmitted by insect vectors in hyperendemic areas, cases tend to cluster in children and adolescents because of early exposure and the accumulation of serotype-specific immunity[9]. However, in settings where transmission is relatively low, as in the locations considered in this study, the burden of symptomatic disease may occur primarily among adults because a large fraction of the adult population remains immunologically naïve[10]. Thus, age-dependent exposure to vectors is a major determinant of the age distribution of cases.

We show in this paper that, in Brazil at least, during specific outbreaks, the ratio between the age distribution of the proportion of officially reported cases and the total number of cases for Aedes transmitted infections such as dengue, chikungunya and zika is **independent** of:

- The size of the outbreak;
- The size of the studied population;
- The period when the outbreak occurs; and
- The geographical region considered.

The above statements are supported by a proper analysis of empirical data as demonstrated in this paper.

In summary, the age-dependent distribution of vector-borne infections emerges from a combination of transmission intensity, immunity acquisition, behavioral exposure, and demographic structure. Analyzing age patterns provides valuable insights into epidemiological processes and supports the design of targeted interventions, such as vaccination programs or vector control strategies focused on the most affected age groups. These issues will be addressed in a future work.

# METHODS

## The available data

To test our hypothesis empirical evidence is required. We used age-dependent incidence of dengue for different geographical regions and years to illustrate our method. The data were obtained from the National System of Compulsory Notification of Infections of the Brazilian Ministry of Health (SINAN)[11], which provides the dengue incidences stratified by ages for each year from 2001 to 2026.

As demonstrated by Figures (5), (7) and (9), which show the outbreaks of dengue in Brazil as a whole, and in cities of Rio de Janeiro and Sao Jose do Rio Preto, respectively, dengue is far from a steady state. The number of cases stratified by age is provided in SINAN[11] and is illustrated in table I for the city Sao Jose do Rio Preto, State of Sao Paulo, South Eastern Brazil, in 2019.

| Table I. Age Dependence of the Number of Cases in a Dengue Outbreak in Sao Jose do Rio Preto, 2019 | |
|---|---|
| **Age** | **Number of Cases** |
| <1 | 186 |
| 1-4 | 568 |
| 5-9 | 1573 |
| 10-14 | 2321 |
| 15-19 | 2701 |
| 20-39 | 12213 |
| 40-59 | 9359 |
| 60-64 | 1622 |
| 65-69 | 1106 |
| 70-79 | 1163 |
| >80 | 422 |
| Total | 33237 |

To calculate the age density of cases, we introduced the binning procedure for density estimation shown in Figure 1 (see[12] and Appendix I for details). For this purpose, we considered the midpoints of each age interval[13]. The number of cases in each interval was divided by the width of the age interval (in years).

For example, in the age interval 20-39 years, the number of reported cases shown in table 1, was 12,213. This corresponds to a density of 12,213 divided by 20, that is, 610.65. As mentioned above, reader unfamiliar with these procedures, are referred to Appendix I for detailed explanation.

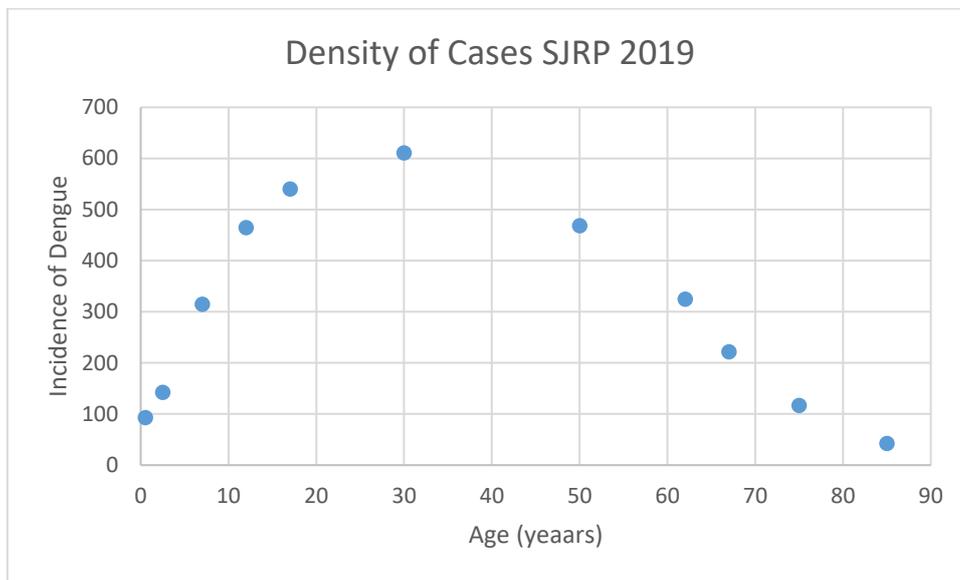

**Figure 1. Age density of dengue cases for the 2019 outbreak in Sao Jose do Rio Preto.**

We fitted a density function, a Gaussian Mesa Function[14], to the data so that we obtain the yearly density function as shown in Figure 2 (see Appendix II for details).

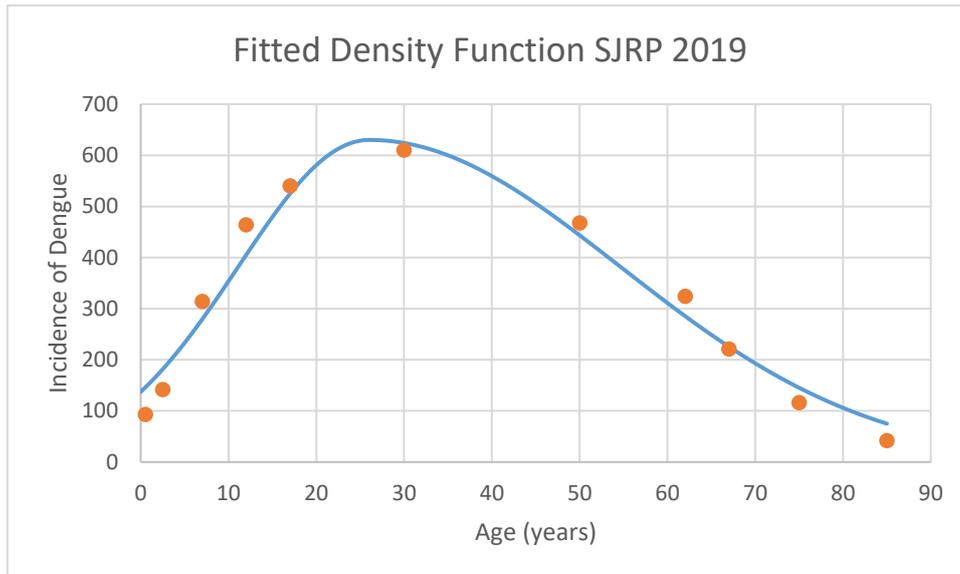

**Figure 2. Fitted age-dependent density function to the data as in Figure 1.**

We used the same procedure for all years outbreaks from 2004-2024 for the three geographical regions analyzed, that is, Brazil as a whole (population 203,080,756), Rio de Janeiro (population 6,211,223) and Sao Jose do Rio Preto (population 480,393) (Figure 3).

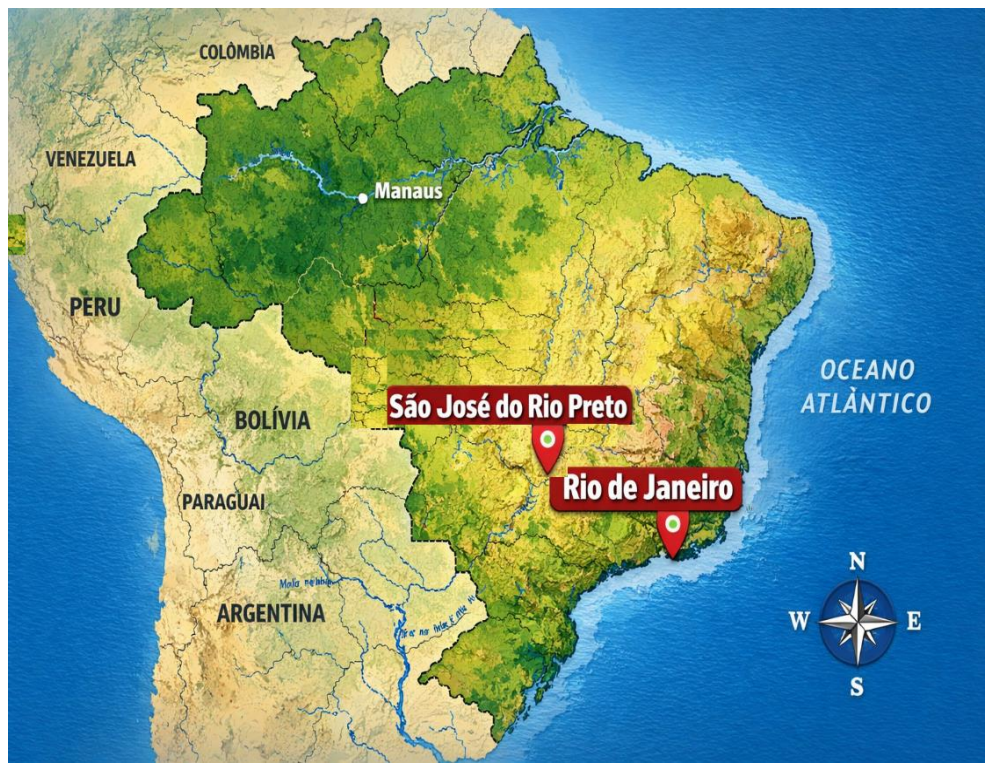

**Figure 3. Map of the locations studied.**

Next, we calculated the *age density proportion of cases*, dividing each point of the above curve by the total number of cases. The results is shown below in Figures (4), (6) and (8), for each site for all years.

We also applied the above procedures to Chikungunya Fever, Zika Fever, Malaria and Visceral Leishmaniosis for Brazil as a whole in the year 2017 (see Figure 11). The same method was applied to Chikungunya and Zika for other cities and other periods and the results are the same.

**Preliminary mathematical considerations**

Let us denote $X(a, Y)$ the double density of new cases of individuals with ages between $a$ and $a + da$ occurring between the times $Y$ and $Y + dY$. The total number of cases at a specific year $Y$, summed over all ages $\Lambda(Y)$, is given by:

$$\Lambda(Y) = \int_0^\infty X(a, Y) da \qquad (1)$$

The right-hand side of equation (1) is the sum of incident cases occurring in a specific year $Y$ across all ages.

Rearranging equation (1), we obtain

$$X(a, Y) = \Lambda(Y) \frac{X(a,Y)}{\int_0^\infty X(a,Y) da} \qquad (2)$$

The fraction $\frac{X(a,Y)}{\int_0^\infty X(a,Y) da}$ in equation (2) represents the proportion of cases at time $Y$, occurring in individuals aged between $a$ and $a + da$. We observed the previously overlooked fact, that this quantity is approximately a function of age $a$ only, independently of both location and the year (see Figures (4), (6) and (8)):

$$\frac{X(a,Y)}{\int_0^\infty X(a,Y)da} = \phi(a) \quad (3)$$

It is remarkable that, for the sites and the years analyzed, this function is independent of both the magnitude of the outbreak and its geographical location, whether within a municipality or across distant regions of the country (see Figure (4), (6) and (8) below). Figure 10 shows the average distribution of the proportion cases across the three regions analyzed, demonstrating their similarity. This observation supports our hypothesis that the distribution depends primarily on the age-dependent exposure of humans to mosquitoes' bites. Furthermore, as shown below, the age distribution of proportion of cases is similar for infections transmitted by aedes mosquitoes and is very different for other vectors (see Figure 11). For dengue, at least, we show below that there is a small difference between the distribution in the proportion of cases between men and women, with the latter showing a slightly larger average age of acquiring the infection. This is to be expected because the behavior of each sex with respect to the exposure to the mosquitoes' bites is probably different (Figure (12) to (14)).

Finally, the average age of prime infection, $\bar{A}$ is given by:

$$\bar{A} = \int_0^L a\phi(a)da \quad (4)$$

where $L$ is the average lifespan of the population (sometimes referred to as 'life expectancy')

**RESULTS**

As mentioned above, Figure (4) shows the age distribution of the proportion of dengue cases in each year in Brazil for the outbreaks occurring from 2014 to 2024.

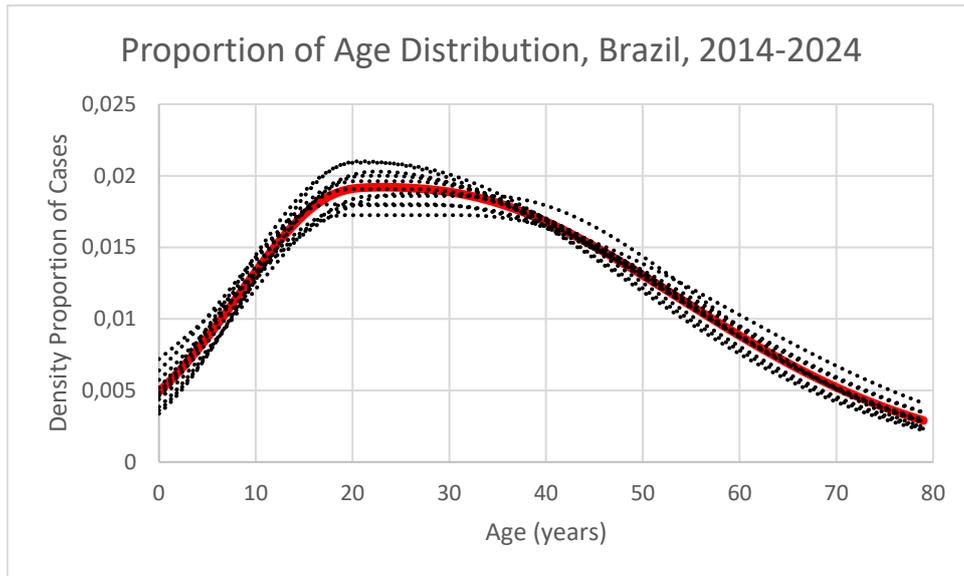

**Figure 4. Age distribution of the proportion of dengue cases for the 2014-2024 outbreaks in Brazil as a whole. Red line represents the average.**

Note that there is a small variation in the distribution of cases with age, although the outbreaks varied considerably from year to year, as shown in Figure 5. The average age (see below) at which individuals acquire dengue in Brazil, for the period analyzed, follows from the average of the curves in Figure 4 and results in 34 years.

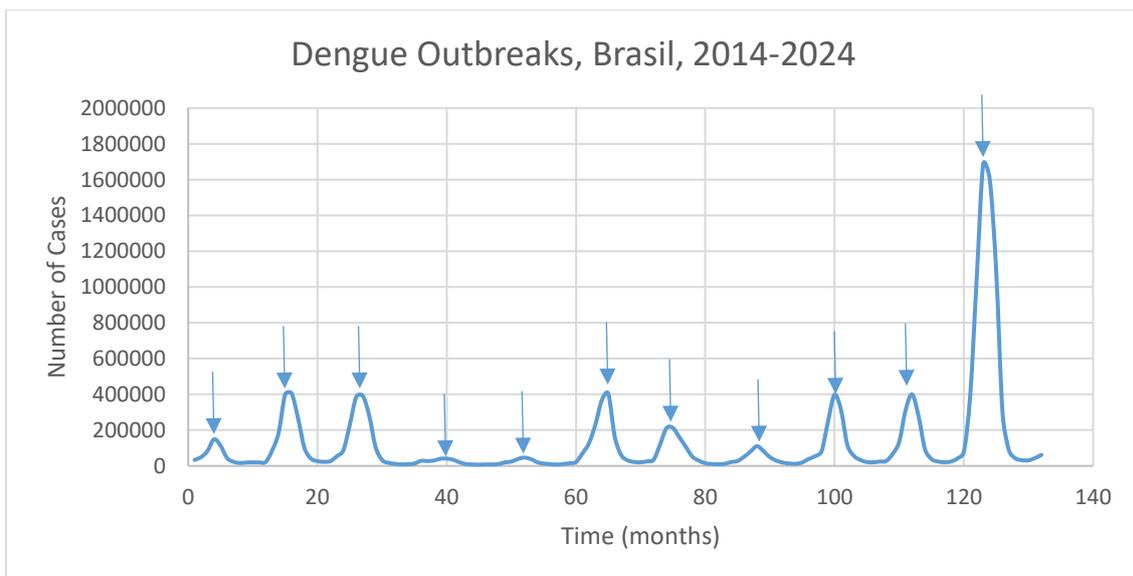

**Figure 5. Dengue outbreaks of dengue in Brazil as a whole for the period 2014-2014. Small arrows indicate the peak of each yearly outbreak,**

Figure 6 shows the age distribution of the proportion of dengue cases in each year in Rio de Janeiro city (RJ) for the outbreaks occurring from 2014 to 2024.

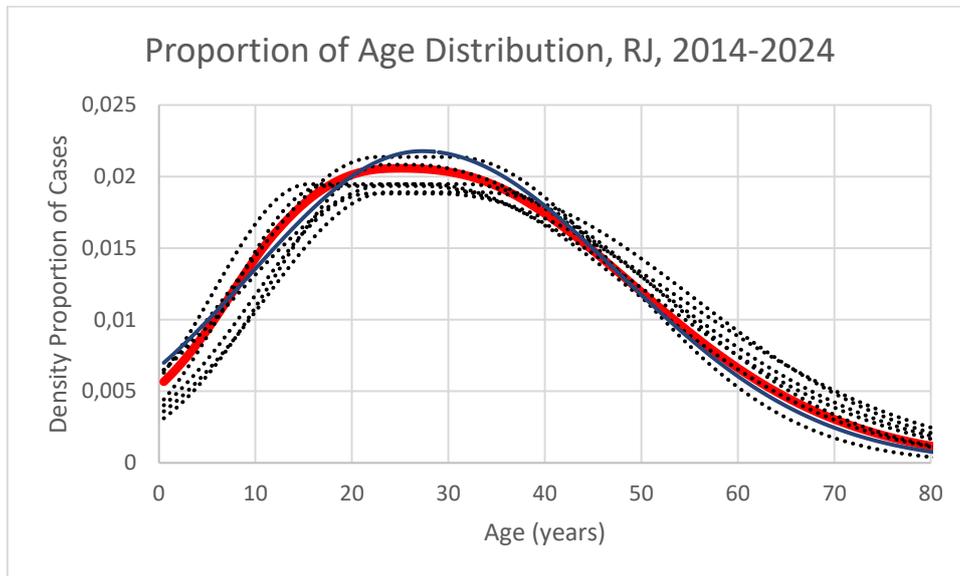

**Figure 6. Age distribution of the proportion of dengue cases for the 2014-2024 outbreaks in Rio de Janeiro city. Red line represents the average.**

Note again, that there is a small variation in the age distribution of cases, independently of the vastly different sizes of the outbreaks in that period, as shown in Figure 7.

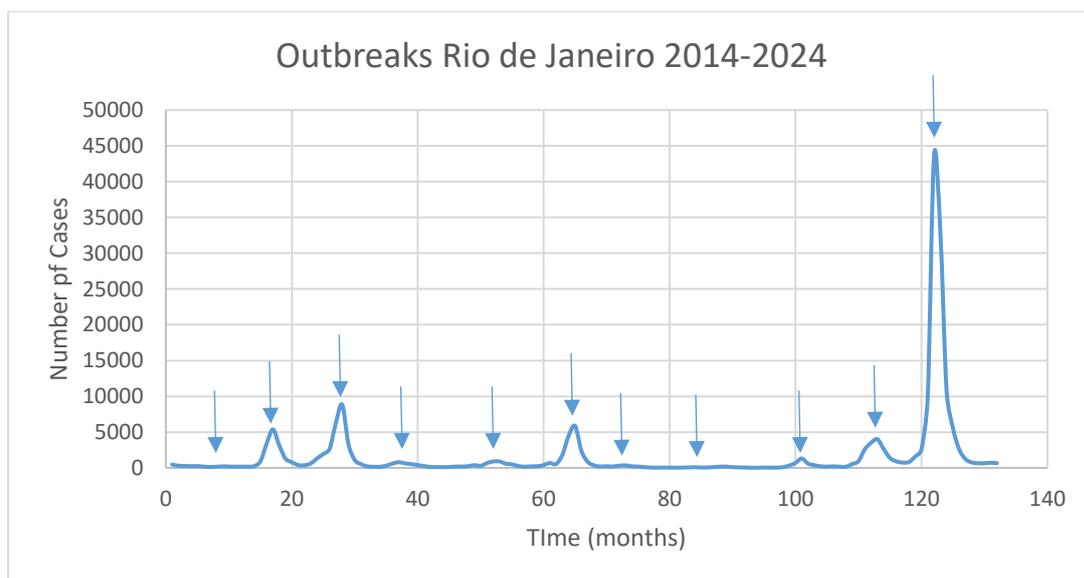

**Figure 7. Dengue outbreaks of dengue in Rio de Janeiro city for the period 2014-2024. Small arrows indicate the peak of each yearly outbreak,**

Figure 8 shows the age distribution of the proportion of dengue cases in each year in Sao Jose do Rio Preto city (SJRP) for the outbreaks occurring from 2014 to 2024.

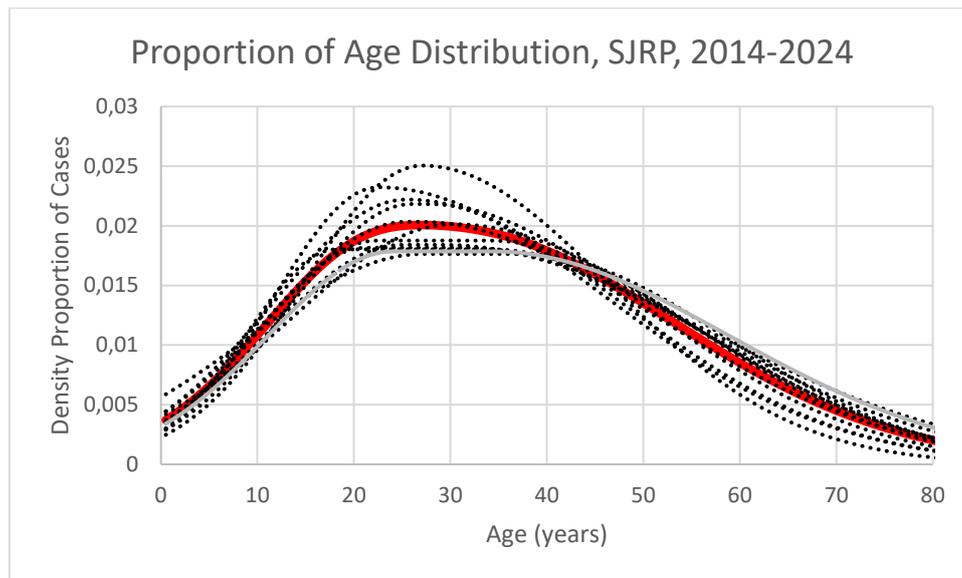

**Figure 8. Age distribution of the proportion of dengue cases for the 2014-2024 outbreaks in Sao Jose do Rio Preto city. Red line represents the average.**

There is a slightly larger variation in the age distribution of proportion of cases (probably due to stochastic variations due to the smaller size of this site) in spite of the great variability in the size of the outbreaks, as shown in Figure 9.

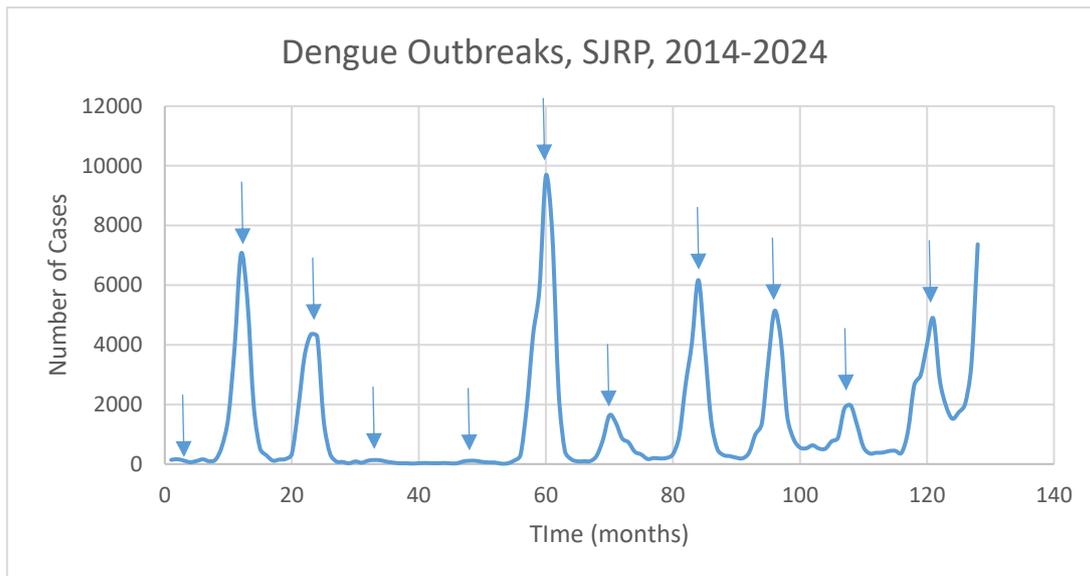

**Figure 9. Dengue outbreaks of dengue in Sao Jose do Rio Preto city for the period 2014-2024. Small arrows indicate the peak of each yearly outbreak.**

The outbreaks in this city are slightly out of phase with respect to the Figures (5) and (7). This is probably due to different ecological interactions between vectors and hosts in this city.

Figure 10 shows the average distribution over the period analyzed of proportion of cases for Brazil (Br), Rio de Janeiro (RJ) and Sao Jose do Rio Preto (SJRP).

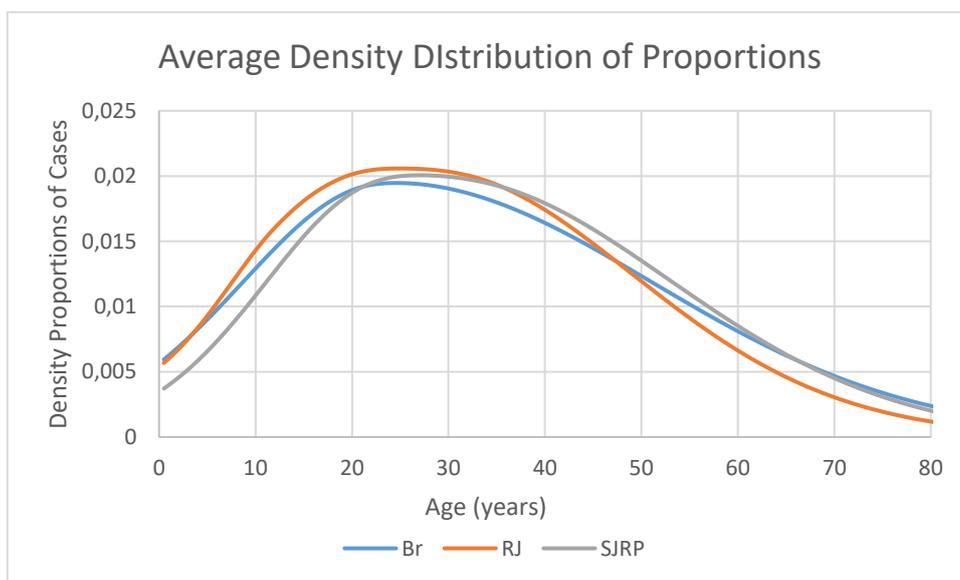

**Figure 10. Average age distribution of the proportion of dengue cases in Brazil as a whole, Rio de Janeiro city and Sao Jose do Rio Preto city.**

It is remarkable that the age distribution of cases are similar for the three geographical areas studied.

In Figure 11 we show the age distribution of the proportion of cases for Dengue, Chikungunya Fever, Zika Fever, Malaria and Visceral Leishmaniosis for Brazil as a whole in the year 2017.

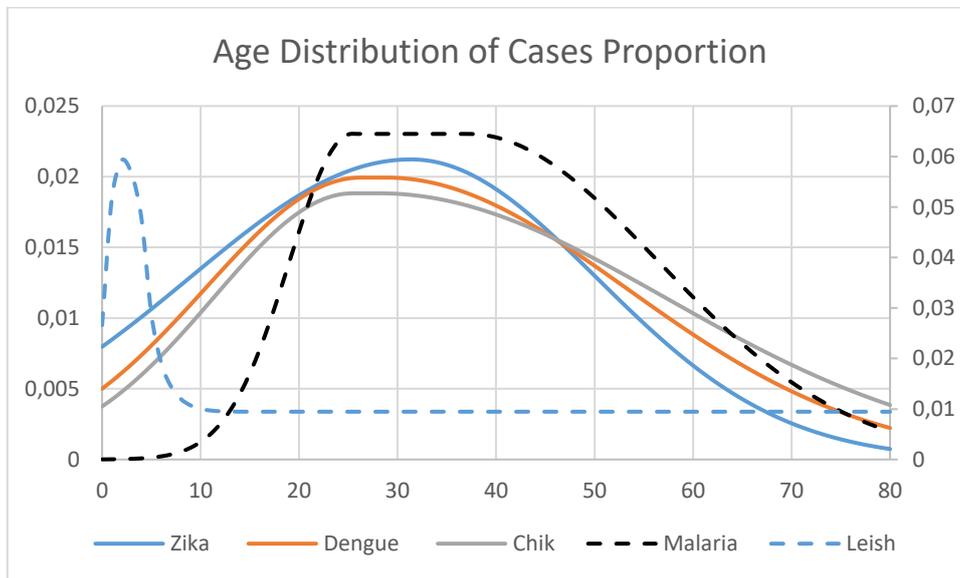

**Figure 11. Distribution of age-dependent proportion of dengue, chikungunya, zika fever, malaria and visceral leishmaniosis cases in Brazil as a whole.**

Note the similarity between the three *Aedes aegypti* transmitted infections, which differ greatly from malaria and visceral leishmaniosis because the two latter infections are transmitted by different vectors[15,16] with different human age exposition to each of them. We are aware that urban yellow fever is also transmitted by *Aedes aegypti* in Africa[17] but in Brazil, it is only transmitted by sylvatic vectors[18] and therefore will not be analyzed in this paper.

Next, we stratified the proportion of cases of dengue in 2024 by sex for the three geographical regions considered, and used the 2024 outbreak as an illustration.

Figure 12 shows the sex-related results for Brazil as a whole. The number (size of the outbreak) of reported cases in 2024 were 2,912,187 and 3,504,989 dengue cases for men and women, respectively.

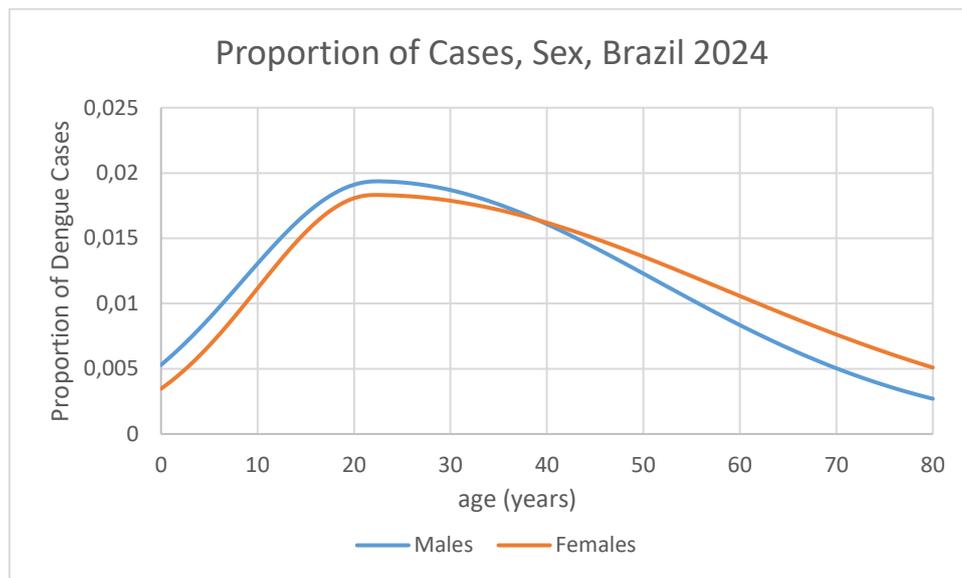

**Figure 12. Distribution of age-dependent proportion of dengue cases for men and women in Brazil as a whole.**

The average age of infection was 33 years and 37 years for men and women, respectively.

Figure 13 shows the results for the city of Rio de Janeiro. The number of reported cases of dengue (size of the outbreak) were 48,116 and 61,822 cases for men and women, respectively.

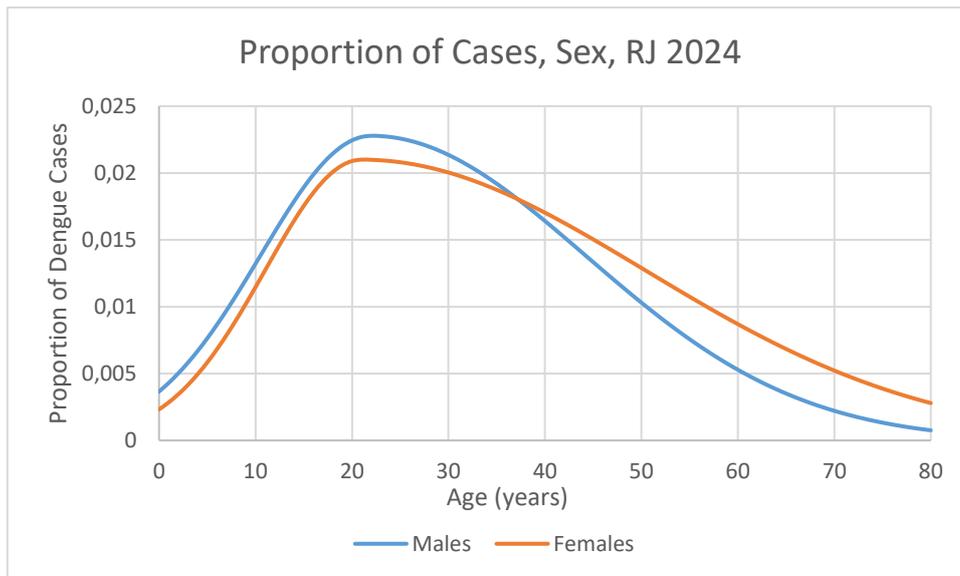

**Figure 13. Distribution of age-dependent proportion of dengue cases for men and women in Rio de Janeiro city.**

The average age of infection was 29 years and 35 years for men and women, respectively.

Figure 14 shows the results for the city of Sao Jose do Rio Preto. The number of reported dengue cases were (size of the outbreak) 16,612 and 19,703 cases for men and women, respectively.

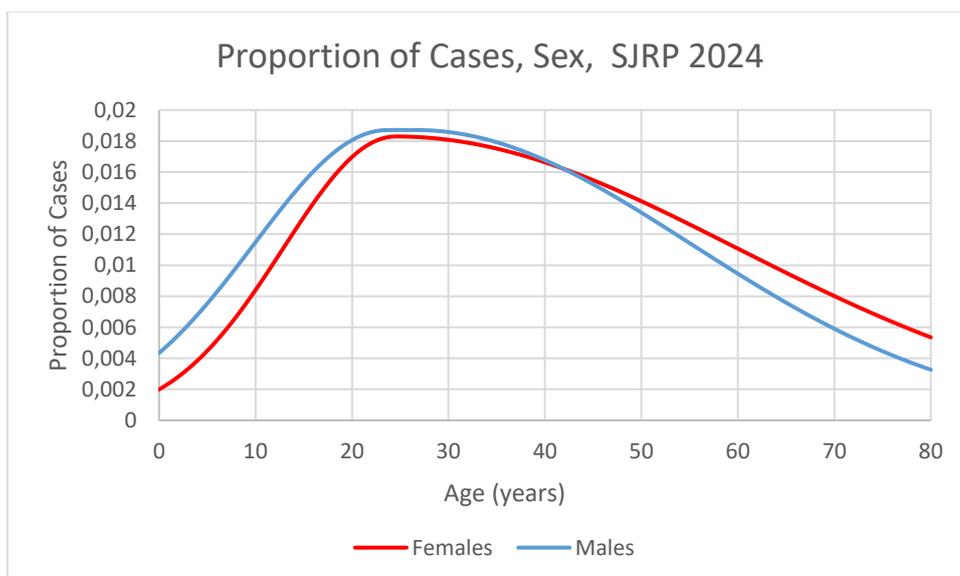

**Figure 14. Distribution of age-dependent proportion of dengue cases for men and women in Sao Jose do Rio Preto city.**

The average age of infection was 35 years and 38 years for men and women, respectively.

Note that the differences between sexes are similar in the three populations analyzed.

We think that the small differences observed between sexes do not justify separate control strategies for each sex to be considered in a future work.

**DISCUSSION**

In this paper, we propose a new approach to analyze the empirical observation that the age distribution of the proportion of reported cases for infections transmitted by Aedes mosquitoes. We found that these distributions are approximately independent of transmission intensity and geographical area. In fact, the age distribution of the proportion of cases is remarkably similar for the three main infections transmitted by *Aedes aegypti* - dengue, chikungunya and Zika, in Brazilian urban centers.

Because the distribution depends mainly on the exposure of hosts to mosquito's bites, it differs for infections transmitted by other vectors (Figure 11) and may differ in other countries where this exposure patterns may vary.

As mentioned above, using the officially reported number of cases for three geographical regions, we found that the age distribution of cases does not differ significantly across sites or across years. This finding may be used in designing age-target interventions and optimizing disease-control strategies, which will be explored in a future work.

As shown in the literature[8], outbreaks in cities occur in different neighborhoods and travels along to other areas of the city. Therefore, our results for a given city assumes that each outbreak occurs in a region where there is a considerable fraction of susceptible individuals. Note also that dengue has at least four different serotypes, differences that were not considered in our analysis.

Some important limitations of our approach are worth mentioning. The first is that by basing the calculations on the officially reported number of cases, the model considers neither underreporting nor recovering from infection. However, underreporting is important only if it modifies the function $\phi(a)$ (see equation (3)), for instance if underreporting occurs predominantly in a certain age interval. This study is valid only for reported cases. However, to use our results to design an intervention, underreporting is important only because it is not necessary to vaccinate everybody in a given age interval since some of them may be already immune by previous infections. The result is that the intervention may be overestimated. Recovering from infection does not modify the model's performance since it would not change the function $\phi(a)$.

Also important to mention is that our results do not allow the proposal of interventions intended to limit the size of the vector population[19,20].

Another limitation of our approach is that, for dengue, there are 4 serotypes. In an ideal scenario, all the above calculations should be done stratifying the cases by serotype. However, this limitation, although important, is valid for dengue only and the same theoretical approach is valid for other Aedes-mediated vector-borne infections. We think that the similarity of the age distribution of cases between the three infections justify not distinguishing serotypes for the case of dengue.

The most important finding of this paper is that the age dependent distribution of the proportion of cases does not vary, neither with the intensity of the transmission nor with the geographical area in Brazil. This function may be completely different in other countries, depending on the age distribution for which exposition to the mosquitoes' bites occurs, because this may modify the form of $\phi(a)$ and may even on the size of the outbreak (see Equation (3)). For instance, contrasting with the average age of dengue acquisition of 34 years in Brazil, in Thailand, this average is lower, about 24 years[21].

Finally, our approach completely ignores stochasticity. One consequence of its deterministic nature is that $\phi(a)$ remains essentialy unmodified, even if the number of cases is so small that it does not allow infection in all ages. For instance, suppose that in a certain outbreak we have less than eighty cases in total. Then, some ages intervals will have no cases but our model predicts fractions of cases for all ages.

**CONCLUSIONS**

Our results allow us to conclude that, in Brazil at least, during specific outbreaks, the ratio between the age distribution of the proportion of officially reported cases and the total number of cases for Aedes transmitted infections such as dengue, chikungunya and zika is **independent** of:

- The size of the outbreak;
- The size of the studied population;
- The period of the outbreak occurs; and
- The geographical region considered.

Our results also suggest that the age-distribution of cases is mainly due to the interaction between vectors and their hosts. Since the interaction between the vectors and male and female hosts may be different, we investigate this and, in fact, the age-distribution is slightly different.

# Appendix I

In this appendix, we explain that treatment of the original data from SINAN[11]. The binning of SINAN[11] is reproduced in Table I of the main text, which has unequal age intervals.

Our problem is how fit a continuous curve to these data. A first, and quite natural but wrong way, of doing it is to plot the raw data from Table I, with age as abscissa (the x-axis) and the number of cases of each age interval in the middle of the interval, as shown in Figure AI.

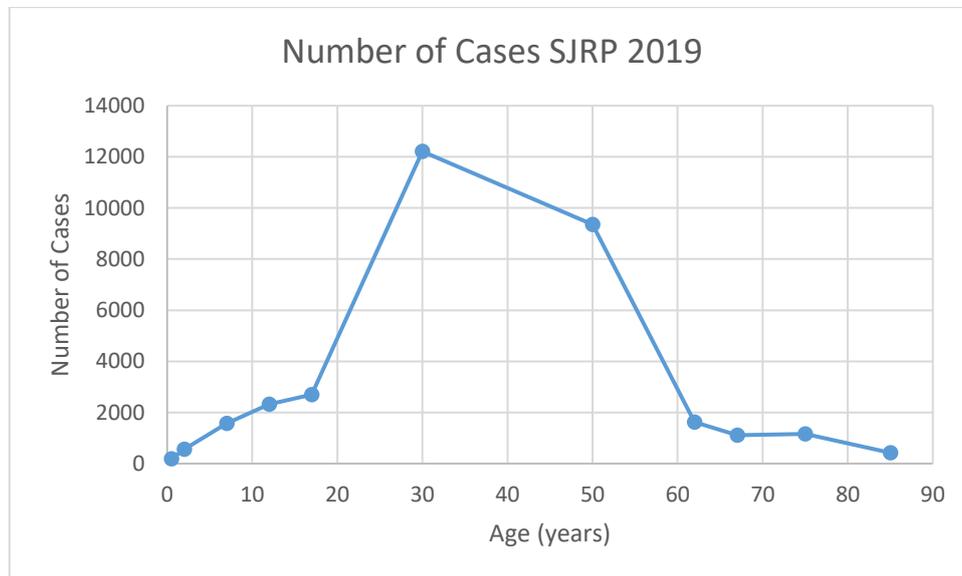

**Figure AI. Plot of the original number of cases against age.**

This is actually misleading and quite meaningless. The graphs seems to indicate that, for instance, at age 20 we have about 5,000 cases, at age 25 we have about 7,000 cases, at age 30 we have about 12,000 cases and at age 35 we have about 11,000. Only considering 4 ages, we have about 34,000 cases, which exceeds the total number of observed cases.

However, if the ordinate (the y-axis) is chosen as *density of cases* rather than *number of cases*, this inconsistency disappears. By density, we mean the number of cases in each age interval divided by the length of the interval.

The right table now reads:

| Table AI. Age Dependence of the Density of Cases in a Dengue Outbreak in Sao Jose do Rio Preto, 2019 | |
|---|---|
| **Age** | **Density of Cases** |
| <1 | 93 |
| 1-4 | 142 |
| 5-9 | 314.6 |
| 10-14 | 464.2 |
| 15-19 | 540.2 |
| 20-39 | 610.65 |
| 40-59 | 467.95 |
| 60-64 | 324.4 |
| 65-69 | 221.2 |
| 70-79 | 116.3 |
| >80 | 42.2 |

Figure AII (Figure 1 of the main text) shows the density plot. The ordinate now is the derivative of the number of cases with respect to ages.

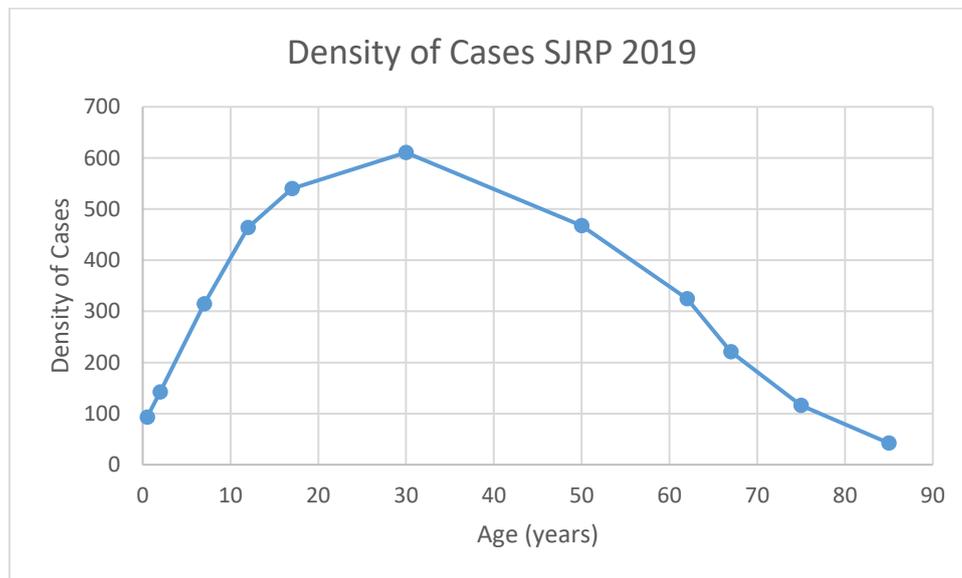

**Figure AII. Plot of density of cases with respect to ages.**

Each point in this graph, say at age $a$, represents the number of cases in a vanishingly small interval between $a$ and $a + da$, so that the total number of cases is obtained by taking the integral of the curve, that is, the area below the curve. This can be checked by integrating the Gaussian Mesa Function[14] that fits the density (see Appendix II below), as in figure 2 of the main text.

# Appendix II

In the main text, and in the Appendix I, we mentioned that we used a Gaussian Mesa Function[14] to fit the density of cases and the distribution of the proportion of cases with respect to age.

In this appendix, we briefly explain what is this function.

The Gaussian mesa function (also called a *flat-top Gaussian* or *mesa beam profile*) is a modification of the classical Gaussian function designed to produce a central plateau with smooth Gaussian edges. Such functions are widely used in optics, laser physics, and beam-shaping applications where a nearly uniform intensity distribution is desired while preserving smooth, differentiable boundaries.

The shape and the analytic form of this function is given below.

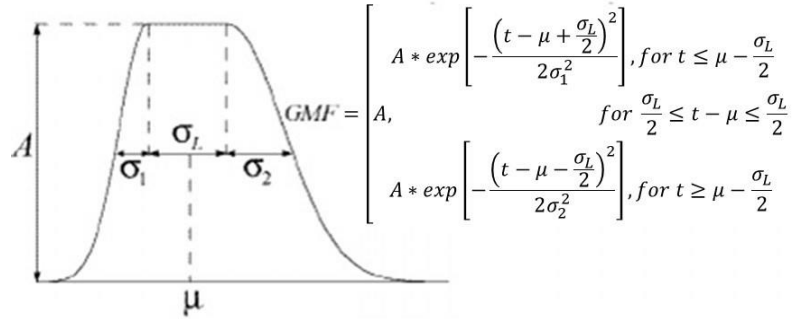

$$GMF = \begin{cases} A * \exp\left[-\frac{\left(t - \mu + \frac{\sigma_L}{2}\right)^2}{2\sigma_1^2}\right], & \text{for } t \leq \mu - \frac{\sigma_L}{2} \\ A, & \text{for } \frac{\sigma_L}{2} \leq t - \mu \leq \frac{\sigma_L}{2} \\ A * \exp\left[-\frac{\left(t - \mu - \frac{\sigma_L}{2}\right)^2}{2\sigma_2^2}\right], & \text{for } t \geq \mu - \frac{\sigma_L}{2} \end{cases}$$

As shown in the figure, parameter $A$ represents the maximum density, which occurs in the interval $\mu - \frac{\sigma_L}{2} \leq t \leq \mu + \frac{\sigma_L}{2}$, $\sigma_1$ is half the width at $\frac{A}{2}$ of Gaussian on the left, $\sigma_2$ is half the width at $\frac{A}{2}$ of Gaussian on the right and $\mu$ is the abscissa at half the plateau.

As an example, the parameters that best fit the data in Figure 2 of the main text are: $A = 630\ cases/year$; $\sigma_L = 0.042\ years$; $\sigma_1 = 14.9\ years$; $\sigma_2 = 28.6\ years$ and $\mu = 26\ years$.

The reader can check that

$$\int_0^{\mu-\sigma_L/2} A\exp\left[-\frac{(t-\mu-\sigma_L/2)^2}{2\sigma_1^2}\right]dt + \int_{\mu-\sigma_L/2}^{\mu+\sigma_L/2} A\,dt + \int_{\mu+\sigma_L/2}^{L} A\exp\left[-\frac{(t-\mu+\sigma_L/2)^2}{2\sigma_2^2}\right]$$
$$= 33{,}325\ cases$$

where, as mentioned in the main text (Equation (4)), $L$ is the average lifespan of the population (sometimes referred to as 'life expectancy').

**References**

1. Botosso V, Precioso AR, Wilder-Smith A, Oliveira D, Oliveira F, Oliveira C, Soares C, Oliveira L, Santos R, Utescher C, Coutinho F, Massad E. Seroprevalence of Zika in Brazil stratified by age and geographic distribution. Epidemiol Infect. 2023 Nov 15;151:1-16. doi: 10.1017/S0950268823001814. Epub ahead of print. PMID: 37965751; PMCID: PMC10728971.
2. Massad E, Burattini MN, Azevedo Neto RS, Yang HM, Coutinho FAB and Zanetta DMT (1994). A model-based design of a vaccination strategy against rubella in a non-immunized community of São Paulo, Brazil. *Epidemiology & Infection* 112: 579-594.
3. Braga C, Luna CF, Martelli CMT, de Souza WV, Cordeiro MT, Alexander N et al. Seroprevalence and risk factors for dengue infection in socio-economically distinct areas of Recife, Brazil. Acta Tropica. 2010 Mar;113(3):234-40. 11.
4. Amaku M, Coudeville L, Massad E. Designing a vaccination strategy against dengue. Rev Inst Med Trop Sao Paulo. 2012 Oct; 54 Suppl 18:S18-21.
5. Reiner, R. C., Perkins, T. A., Barker, C. M., et al. (2013). A systematic review of mathematical models of mosquito-borne pathogen transmission. *Journal of the Royal Society Interface*, 10(81), 20120921.
6. Guzman, M. and Kouri, G., 2002. Dengue: an update. *The Lancet Infectious Diseases*, vol. 2, no. 11, p. 33-42.
7. Cecílio, ABI, Campanelli, ES, Souza, KPR, Figueiredo, IB, Resende, MC (2009). Natural vertical transmission by *Stegomyia albopicta* as dengue vector in Brazil. Braz. J. Biol. 69 (1) doi 10.1590/S1519-69842009000100015.
8. Amaku M, Azevedo F, Burattini MN, Coelho GE, Coutinho FAB, Greenhalgh D, Lopez LF, Motitsuki RS, Wilder-Smith A, Massad E (2016). Magnitude and frequency variations of vector-borne infection outbreaks using the Ross-Macdonald model: explaining and predicting outbreaks of dengue fever. *Epidemiology & Infection.* 144 (16):3435-3450.
9. Guzman, MG, Halstead, SB, Artsob, H, et al. (2010). Dengue: a continuing global threat. *Nature Reviews Microbiology*, 8(12), S7–S16.
10. Halstead, SB. (2007). Dengue. *The Lancet*, 370(9599), 1644–1652.

11. SINAN-MOH (2025). Sistema Nacional de Agravos de Notificação. Ministério da Saúde do Brasil. https://datasus.saude.gov.br/acesso-a-informacao/doencas-e-agravos-de-notificacao-de-2007-em-diante-sinan/ Accessed in 2 January 2025.

12. Millman J and Seely S (1951). Eletronics. New York, McGraw-Hill Book Company.

13. Cole Notes (2009). Statistics & Data. Toronto. Coles Publishing.

14. Coutinho FAB, Amaku M, Boulos FC, de Sousa Moreira JA, de Barros ENC, Kallas EG, Massad E (2025). Estimating the Size of the Aedes Mosquitoes' Population Involved in Outbreaks of Dengue and Chikungunya Using a Mathematical Model. *Bulletin of Mathematical Biology.* 87 (9):124.

15. Oliveira TMP, Laporta GZ, Bergo ES, Chaves LSM, Antunes JLF, Bisckersmith AS, Conn JE, Massad E, Sallum MAM (2021. Vector role and human biting activity of Anophelinae mosquitoes in different landscapes in the Brazilian Amazon. *Parasites and Vectors* 14(1):236. doi: 10.1186/s13071-021-04725-2.

16. Shimozako HJ, Wu J, Massad E (2017). Mathematical modelling for Zoonotic Visceral Leishmaniasis dynamics: A new analysis considering updated parameters and notified human Brazilian data. *Infectious Diseases Modelling.* 2 (2):143-160.

17. Wilder-Smith A, Massad E (2018). Estimating the number of unvaccinated Chinese workers against yellow fever in Angola. *BMC Infectious Diseases.* 18 (1):185.

18. Cano ME, Marti GA, Balsalobre A, Muttis E, Bruno EA, Rossi G, Micieli MV. Database of Sabethes and Haemagogus (Diptera: Culicidae) in Argentina: Sylvatic Vectors of the Yellow Fever Virus. *Journal of Medical Entomology*.16;58(4):1762-1770. doi: 10.1093/jme/tjab059.

19. Massad E, Coutinho FA, Lopez LF, da Silva DR (2011). Modeling the impact of global warming on vector-borne infections. *Physics of Life Reviews*. 8 (2):169-99.

20. Ferreira CP, Lyra SP, Azevedo F, Greenhalgh D, Massad E (2017). Modelling the impact of the long-term use of insecticide-treated bed nets on Anopheles mosquito biting time. *Malaria Journal.* 16 (1):373.

21. Huang AT, Takahashi S, Salje H and Cummings DAT. (2022). Assessing the role of multiple mechanisms increasing the age of dengue cases in Thailand. *Proceeding of the National Academy of Sciences USA.* 119 (20) e2115790119.

**Funding.**. This work was partially funded by CNPq and Fundacao Butantan

**Conflict of Interests.** The authors declare no conflict of interest.